\documentclass[12pt]{article}
\usepackage{color}
\usepackage{graphicx}
\usepackage{slashed}
\usepackage{geometry}
\begin{document}
\title{Analytical calculations of ground and excited states of unequal mass heavy pseudoscalar
and vector mesons mass spectra using Bethe-Salpeter formalism}
\author{Hluf Negash$^{1}$ and Shashank Bhatnagar$^{2}$}
\maketitle
$^{1}$Department of Physics, Samara University,P.O.Box 132, Samara, Ethiopia\\
$^{2}$Department of Physics, University Institute of Sciences,
Chandigarh University, Mohali-140413, India\\

\textbf{Abstract}\\
Considering the fact that, a wide range of variation in masses of
various states of heavy pseudoscalar and vector mesons can be seen
in different models and the experimental data on masses of many of
these states is not yet currently available, in this paper, we
study the analytical calculation of heavy pseudoscalar and vector
mesons mass spectra ( such as $\eta_{c}$, $B_{c}$, $\eta_{b}$,
$J/\psi$, $B^{*}_{c}$ and $\Upsilon$) and their corresponding wave
functions in the framework of Bethe-Salpeter equation under the
covariant instantaneous ansatz. The parameters of the framework
were determined by a fit to the heavy pseudoscalar and vector
mesons mass spectrum of ground states. These input parameters so
fixed were found to give good agreements with data on mass spectra
of ground and excited states of these mesons mass spectra.

\section{Introduction}
For a long time, a number of studies dealing with heavy
pseudoscalar and vector mesons mass spectra at quark level of
compositeness have been carried out to understand the internal
structure and dynamics of hadrons. Theoretically, our knowledge of
hadron physics is mainly based on phenomenological quark
confinement models, as the hadron domain generally falls in the
non-perturbative regime of QCD. The analytical calculation of mass
spectrum and the corresponding wave functions are very important
topic for heavy pseudoscalar and vector mesons. Through these
studies, one may have insight of the heavy mesons and lead to
better understanding of QCD. Thus, a realistic description of
pseudoscalar and vector mesons mass spectra framework at the quark
level of compositeness would be an important element in our
understanding of hadron dynamics and internal structures. A number
of approaches such as Lattice QCD \cite{bali97,burch09}, Chiral
perturbation theory \cite{gasser84}, heavy quark effective theory
\cite{neubert94}, QCD sum rules \cite{shifman79,veli12}, N.R.QCD
\cite{brambilla11,bodwin}, dynamical-equation based approaches
like Schwinger-Dyson equation and Bethe-Salpeter equation (BSE)
\cite{smith69,mitra01,munz,alkofer01,wang10,koll,bhatnagar92}, and
potential models \cite{godfrey85,devlani} have been proposed to
deal with the long distance property of QCD. However, since the
task of calculating hadron structures from QCD itself is very
difficult, as can be seen from various Lattice QCD approaches, one
generally relies on specific models to gain some understanding of
QCD at low energies.

\bigskip
Thus, the mass spectrum and the corresponding wave functions of
the heavy quark bound state system have been studied in the
context of many QCD-motivated models. These model predictions give
a wide range of variations. Relativistic quark models are
important in hadronic physics. A relativistic framework for
analyzing mesons as composite objects is provided by
Bethe-Salpeter Equation (BSE), which is fully rooted in field
theory and is a conventional non-perturbative approach in dealing
with relativistic bound state problems in QCD. It provides a
realistic description for analyzing inner structure of hadrons,
which is also crucial in treating hadronic decays and the decay
constants \cite{hluf15,hluf16,hluf17} and is a direct source of
information on the success of any model.

In this paper we will calculate the mass spectra and the
corresponding wave functions of heavy pseudoscalar and vector
mesons for ground and excited states, such as $1S$, $2S$, $3S$ and
$4S$ for  $\eta_{c}$, $B_{c}$ and $\eta_{b}$ mesons and  $1S$,
$2S$, $1D$, $3S$ $2D$, $4S$ and $3D$ for  $J/\psi$, $B^{*}_{c}$
and $\Upsilon$ mesons which are composites of $c$ and $b$ quarks,
in the framework of BSE under  Covariant Instantaneous Ansatz
(CIA)\cite{bhatnagar14}.  This work is a generalization of our
previous work \cite{hluf16} (involving equal mass quarkonia) to
calculation of mass spectra of unequal mass quarkonia. We have
made use of unequal mass kinematics, using the Wight-mann Garding
definition of masses \cite{mitra01,bhatnagar14}. This effectively
increases the number of number of states of mesons studied. The
correctness of our generalization to unequal mass quarks can be
gauged by the fact that under equal mass condition ($m_1=m_2$),
all our equations reduce to the corresponding equations for equal
mass quarkonia such as, $\eta_c, \eta_b, J/\Psi$, and $\Upsilon$
in \cite{hluf16}.

To derive the mass spectral equation and their solutions, i.e. the
mass spectrum and the wave functions, we have to start with the
most general structure of Bethe-Salpeter wave function for
pseudoscalar ($J^{PC}=0^{-+}$ state) and vector ($J^{PC}=1^{--}$
state) mesons \cite{smith69}. We then put the formulated BS wave
functions into the instantaneous BSE (Salpeter equations) and turn
the equation into a set of proper coupled equations for the
components which appear in the formulation. These equations are
then explicitly shown to decouple in the heavy-quark
approximation, and are reduced to a single mass spectral equation
for any given state, whose analytic solutions in an approximate
harmonic oscillator basis yield the mass spectrum and wave
functions for ground and excited states of pseudoscalar and vector
mesons, $\eta_{c}$, $B_{c}$, $\eta_{b}$, $J/\psi$, $B^{*}_{c}$ and
$\Upsilon$. We have explicitly shown that in the heavy quark
approximation (valid for heavy quarkonia), these equations can be
decoupled, and finally analytical solutions (both mass spectrum
and eigenfunctions) of this equation can be obtained using
approximate harmonic oscillator basis.
\bigskip

Our main focus in this paper was to show that this problem of
$4\times 4$ BSE under heavy quark approximation can indeed be
handled analytically for both masses as well as the wave
functions. With the above approximation, that is quite justifiable
and well under control, in the context of heavy quark systems, we
have been able to give analytical solutions of mass spectral
equations for heavy pseudoscalar and vector mesons, giving us a
much deeper insight into this problem. Analytical forms of the
wave functions for $S$ and $D$ state for $n=0,1,2$ for
$c\overline{c}$, $c\overline{b}$ and $b\overline{b}$ systems thus
obtained. The analytical forms of eigenfunctions for ground and
excited states so obtained will be used to evaluate many
transition amplitudes of these mesons. In this paper, we are
focusing on the analytical calculation of the ground and excited
states of unequal mass heavy pseudoscalar and vector mesons
spectroscopy using Bethe-Salpeter equation and our future study
will be the calculation of transition amplitudes involving these
mesons. We have also plotted the graphs of all the wave functions
for the states, $1S, ... ,2D$. The over all features of all these
plots show that the states $nS$ have $n-1$ nodes in their wave
functions, and are very similar to the corresponding plots of wave
functions in \cite{wang10}, obtained by purely numerical methods,
which show the correctness of our approach.

\bigskip
This paper is organized as follows: In section \textbf{2}, we give
the details of our framework, and the derivation of mass spectral
equations using the BS wave functions of heavy pseudoscalar and
vector mesons. Finally, discussions and conclusion are given in
section \textbf{3}.

\section{Formulation of the mass spectral equation under BSE}
We start from the derivation of Salpeter equations in this
section, giving only the main steps. The 4D BSE for $q\bar{q}$
comprising of quarks-antiquark of momenta $p_{1,2}$, and masses
$m_{1}$ and $m_{2}$ respectively is written in $4\times 4$
representation as:

\begin{equation}
(2\pi)^{4}
\left(\slashed{p}_{1}-m_{1}\right)\Psi(P,q)\left(\slashed{p}_{2}+m_{2}\right)
= i\int d^{4}q'K(q,q')\Psi(P,q')
\end{equation}
where $\Psi(P,q)$ is the 4D BS wave function, $K(q,q')$ is the
interaction kernel between the quark and anti-quark, and $p_{1,2}$
are the momenta of the quark and anti-quark, which are related to
the internal 4-momentum $q$ and total momentum $P$ of hadron of
mass $M$ as,

\begin{equation}
p_{1,2\mu} = \hat{m}_{1,2}P_{\mu} \pm q_{\mu}
\end{equation}

where
\begin{equation}
\hat{m}_{1,2}=\frac{1}{2}\left(1\pm\frac{m^{2}_{1}-m^{2}_{2}}{M^{2}}\right)
\end{equation}
are the Wightman-Garding (WG) \cite{ mitra01,bhatnagar14}
definitions of masses of individual quarks which ensure that
$P.q=0$ on the mass shells of either quarks, even when $m_1\neq
m_2$. These WG definitions of masses also ensure an effective
partitioning of internal momentum for unequal mass quarks in the
hadron. We decompose the internal momentum, $q_\mu$ as the sum of
its transverse component, $\hat{q}_\mu=q_\mu-(q\cdot P)P_\mu/P^2$
(which is orthogonal to total hadron momentum $P_\mu$), and the
longitudinal component, $\sigma P_\mu = (q\cdot P)P_\mu/P^2$,
(which is parallel to $P_\mu$). Thus,
$q_\mu=(M\sigma,\widehat{q})$, where the transverse component,
$\widehat{q}$ is an effective 3D vector, while the longitudinal
component, $M\sigma$ plays the role of the time component. The 4-D
volume element in this decomposition is,
$d^4q=d^3\hat{q}Md\sigma$. To obtain the 3D BSE and the
hadron-quark vertex, use an Ansatz on the BS kernel $K$ in Eq. (1)
which is assumed to depend on the 3D variables $\hat{q}_\mu$,
$\hat{q}_\mu^\prime$ as,
\begin{equation}
 K(q,q') = K(\hat{q},\hat{q}')
\end{equation}
Hence, the longitudinal component, $M\sigma$ of $q_{\mu}$, does
not appear in the form $K(\hat{q},\hat{q}')$ of the kernel and we
define 3D wave function $\psi(\hat{q})$ as:
\begin{equation}
\psi(\hat{q}) = \frac{i}{2\pi}\int Md\sigma\Psi(P,q)
\end{equation}
Substituting Eq.(5) in Eq.(1), with definition of kernel in
Eq.(4), we get a covariant version of Salpeter equation,
\begin{equation}
(2\pi)^{3}
\left(\slashed{p}_{1}-m_{1}\right)\Psi(P,q)\left(\slashed{p}_{2}+m_{2}\right)=
\int d^{3}\hat{q}' K(\hat{q},\hat{q}')\psi(\hat{q}')
\end{equation}

and the 4D BS wave function can be written as,
\begin{equation}
\Psi(P,q) = S_{F}(p_{1})\Gamma(\hat{q})S_{F}(-p_{2})
\end{equation}
where
\begin{equation}
\Gamma(\hat{q})=\int \frac{d^{3}\hat{q}'}{(2\pi)^{3}}
K(\hat{q},\hat{q}')\psi(\hat{q}')
\end{equation}
 plays the role of hadron-quark vertex function. Following a
 sequence of steps given in \cite{hluf15,hluf16},
we obtain four Salpeter equations:
\begin{eqnarray}
 &&\nonumber(M-\omega_1-\omega_2)\psi^{++}(\hat{q})=-\Lambda_{1}^{+}(\hat{q})\Gamma(\hat{q})\Lambda_{2}^{+}(\hat{q})\\&&
   \nonumber(M+2\omega_1+\omega_2)\psi^{--}(\hat{q})=\Lambda_{1}^{-}(\hat{q})\Gamma(\hat{q})\Lambda_{2}^{-}(\hat{q})\\&&
 \psi^{+-}(\hat{q})=\psi^{-+}(\hat{q})=0
\end{eqnarray}
with  the energy projection operators,
$\Lambda_{j}^{\pm}(\hat{q})=\frac{1}{2\omega_{j}}\bigg[\frac{\slashed{P}\omega_{j}}{M}\pm
I(j)(m_{j}+\slashed{\hat{q}})\bigg]$,
$\omega_{j}^{2}=m_{j}^{2}+\hat{q}^{2}$, and $I(j)=(-1)^{j+1}$
where $j=1, 2$ for quarks and anti-quarks respectively. The
projected wave functions, $\psi^{\pm\pm}(\hat{q})$ in Salpeter
equations are obtained by the operation of the above projection
operators on $\psi(\widehat{q})$ (for details see
\cite{hluf15,hluf16} as,
\begin{equation}
 \psi^{\pm\pm}(\hat{q})=\Lambda_{1}^{\pm}(\hat{q})\frac{\slashed{P}}{M}\psi(\hat{q})
 \frac{\slashed{P}}{M}\Lambda_{2}^{\pm}(\hat{q})
\end{equation}

To obtain the mass spectral equation, we have to start with the
above four Salpeter equations and solve the instantaneous Bethe
Salpeter equation. However, the last two equations do not contain
eigenvalue $M$, and are thus employed to obtain constraint
conditions on the Bethe - Salpeter amplitudes associated with
various Dirac structures in $\psi(\widehat{q})$, as shown in
details in \cite{hluf16}. The framework is quite general so far.
In fact the above four equations constitute an eigenvalue problem
that should lead to evaluation of heavy pseudoscalar and vector
mesons mass spectrum, such as $\eta_{c}$, $B_{c}$, $\eta_{b}$,
$J/\psi$, $B^{*}_{c}$ and $\Upsilon$.
\bigskip

We now show the method for calculation of heavy pseudoscalar and
vector mesons mass spectrum. We first write down the most general
formulation of the relativistic BS wave functions, according to
the total angular momentum (J), parity (P) and charge
conjugation(C) of the concerned bound state. We then put this BS
wave function in Eq.(9) to derive the mass spectral equations,
which are a set of coupled equations.

\begin{enumerate}
\item \textbf{For pseudoscalar mesons}, the complete decomposition of 4D BS wave
function in terms of scalar functions $\phi_{i}(P,q)$ is
\cite{smith69,hluf15, hluf16}:
\begin{equation}
\Psi_{P}(P,q) \approx
\left\{\phi_{1}(P,q)+\slashed{P}\phi_{2}(P,q)+\slashed{q}\phi_{3}(P,q)
+[\slashed{P},\slashed{q}]\phi_{4}(P,q)\right\}\gamma_{5}
\end{equation}
Following \cite{wang10,hluf15,hluf16}, we write the general
decomposition of the instantaeous BS wave function for
pseudoscalar mesons $(J^{PC}=0^{-+})$, in the center of mass frame
as:
\begin{equation}
\psi_{P}(\hat{q}) \approx
\left\{M\phi_{1}(\hat{q})+\slashed{P}\phi_{2}(\hat{q})+\slashed{\hat{q}}\phi_{3}(\hat{q})
+\frac{\slashed{P}\slashed{\hat{q}}}{M}\phi_{4}(\hat{q})\right\}\gamma_{5}
\end{equation}
where $\phi_{1}$, $\phi_{2}$, $\phi_{3}$ and $\phi_{4}$ are even
functions of $\hat{q}$ and M is the mass of the bound state. Now
lets come to derive the mass spectral coupled equations from
Eq.(9). Putting Eq.(12) into the last two equations of Eq.(9):
\begin{equation}
\psi^{+-}(\hat{q})=\psi^{-+}(\hat{q})=0
\end{equation}
One can obtain:
\begin{equation}
\phi_{3}=\frac{\phi_{1}M(-\omega_{1}+\omega_{2})}{\omega_{1}m_{2}+\omega_{2}m_{1}}
;
\phi_{4}=-\frac{\phi_{2}M(\omega_{1}+\omega_{2})}{\omega_{1}m_{2}+\omega_{2}m_{1}}
\end{equation}
So we can apply Eq.(14) into Eq.(12) and rewrite the relativistic
wave function of $0^{-+}$ as:
\begin{equation}
\psi_{P}(\hat{q}) \approx
\left\{\left(M+\frac{\slashed{\hat{q}}M(\omega_{2}-\omega_{1})}{m_{2}\omega_{1}+m_{1}\omega_{2}}\right)
\phi_{1}(\hat{q})+\left(\slashed{P}\phi_{2}(\hat{q})+\frac{\slashed{\hat{q}}\slashed{P}(\omega_{2}+\omega_{1})}{m_{2}\omega_{1}+m_{1}\omega_{2}}
\right)\phi_{2}(\hat{q})\right\}\gamma_{5}
\end{equation}
where, we have been able to express the instantaneous wave
function $\psi^{P}(\widehat{q})$ in terms of only the leading
Dirac structures associated with amplitudes $\phi_{1}$ and
$\phi_{2}$. Put the wave function Eq.(15) into the first two
Salpeter equations of Eq.(9) and by evaluating traces of
$\gamma$-matrices on both sides, we obtain the independent BS
coupled integral equations:
\begin{eqnarray}
 &&\nonumber (M-\omega_{1}-\omega_{2})[H_{1}\phi_{1}
+H_{2}\phi_{2}]=\int\frac{d^{3}\hat{q}'}{(2\pi)^{3}}K(\hat{q},\hat{q}')
 \left[H'_{1}\phi'_{1}+H'_{2}\phi'_{2}\right]\\&&
(M+\omega_{1}+\omega_{2})[H_{1}\phi_{1}-H_{2}\phi_{2}]
 =-\int\frac{d^{3}\hat{q}'}{(2\pi)^{3}}K(\hat{q},\hat{q}')
 \left[H'_{1}\phi'_{1} -H'_{2}\phi'_{2}\right]
\end{eqnarray}
where
\begin{eqnarray}
 &&\nonumber H_{1}=M(\omega_{1}\omega_{2}+m_{1}m_{2}- \hat{q}^{2})+A(m_{1}-m_{2})\hat{q}^{2}\\&&
    \nonumber H_{2}=M\left[\omega_{1}m_{2}+m_{1}\omega_{2}-B(\omega_{1}+\omega_{2})\hat{q}^{2}\right]\\&&
    \nonumber H'_{1}=M(\omega_{1}\omega_{2}+m_{1}m_{2}- \hat{q}^{2})-A'(m_{1}-m_{2})\hat{q}.\hat{q}'\\&&
  \nonumber H'_{2}=M\left[\omega_{1}m_{2}+m_{1}\omega_{2}-B'(\omega_{1}+\omega_{2})\hat{q}.\hat{q}'\right]\\&&
  \nonumber A=\frac{M(\omega_{2}-\omega_{1})}{\omega_{1}m_{2}+\omega_{2}m_{1}}\\&&
  \nonumber B=\frac{\omega_{1}+\omega_{2}}{\omega_{1}m_{2}+\omega_{2}m_{1}}\\&&
  \nonumber A'=\frac{M(\omega'_{2}-\omega'_{1})}{\omega'_{1}m_{2}+\omega'_{2}m_{1}}\\&&
  \nonumber B'=\frac{\omega'_{1}+\omega'_{2}}{\omega'_{1}m_{2}+\omega'_{2}m_{1}}\\&&
  \nonumber \omega_{1,2}^{2} = m_{1,2}^{2} + \hat{q}^{2}\\&&
  \omega'^{2}_{1,2}= m_{1,2}^{2} + \hat{q}'^{2}
\end{eqnarray}
The correctness of these equations can be gauged by the fact that
under equal mass condition ($m_1=m_2$), these equations reduce to
the corresponding equal mass equations, Eq.(21) for equal mass
pseudoscalar quarkonia $\eta_c$,and $\eta_b$ in \cite{hluf16}.

\item \textbf{For vector mesons}, the complete decomposition of 4D BS wave function in terms of
various Dirac structures is \cite{smith69,hluf16}:

\begin{eqnarray}
&&\nonumber\
\Psi_{V}(P,q)\approx\slashed{\epsilon}\chi_{1}+\slashed{\epsilon}\slashed{P}
\chi_{2} +[q\cdot\epsilon-\slashed{\epsilon}\slashed{
q}]\chi_{3}+[2q.\epsilon\slashed{P}+\slashed{\epsilon}(\slashed{P}\slashed{q}-\slashed{q}\slashed{P})]\chi_{4}\\&&
\
(q.\epsilon)\chi_{5}+(q.\epsilon)\slashed{P}\chi_{6}+(q.\epsilon)\slashed{q}\chi_{7}+
(q.\epsilon)(\slashed{P}\slashed{q}-\slashed{q}\slashed{P})\chi_{8}
\end{eqnarray}
where
$\chi_{\alpha}=\chi_{\alpha}(q^{2},q.P,P^{2});(\alpha=1,2,3,...,8)
$ are the Lorentz scalar amplitudes multiplying the various Dirac
structures in the BS wave function, $\Psi(P,q)$. We can write the
instantaneous BS wave function $\psi^{V}(\hat{q})$ with
dimensionality $M$ for vector quarkonia, $(J^{PC}=1^{--})$ up to
sub-leading order as per the power counting scheme
\cite{bhatnagar14} as in case of pseudoscalar mesons, in the
center of mass frame as \cite{wang10,hluf16}:
\begin{eqnarray}
&&\nonumber\ \psi_{V}(\hat{q})\approx
M\slashed{\epsilon}\chi_{1}+\slashed{\epsilon}\slashed{P}
\chi_{2}+[\slashed{\epsilon}\slashed{\hat{q}}-\hat{q}.\epsilon]\chi_{3}
+[\slashed{P}\slashed{\epsilon}\slashed{\hat{q}}-(\hat{q}.\epsilon)\slashed{P}]\frac{1}{M}\chi_{4}+(\hat{q}.\epsilon)\chi_{5}\\&&
 +(\hat{q}.\epsilon)\slashed{P}\frac{\chi_{6}}{M}
\end{eqnarray}
where $\chi_{1}, ... ,\chi_{6}$ are even functions of $\hat{q}$
and $M$ is the mass of the bound state (of the corresponding
meson). We now derive the mass spectral coupled equations from
Eq.(9). Putting Eq.(19) into the last two equations of Eq.(9), and
we obtain the independent constraints on the components for the
Instantaneous BS wave function:
\begin{equation}
\chi_{5}=\frac{M(\omega_{1}+\omega_{2})\chi_{1}}{(m_{2}\omega_{1}+m_{1}\omega_{2})};
\chi_{4}=-\frac{M(\omega_{1}+\omega_{2})\chi_{2}}{(m_{2}\omega_{1}+m_{1}\omega_{2})}
 \end{equation}
with the heavy quark approximation, $\widehat{q}\ll P$, the terms
$\chi_{6}$ and $\chi_{3}$ have minimum contributions. Then,
applying the constraints in Eq.(20) to Eq.(19), we can rewrite the
relativistic wave function of state $(J^{PC}=1^{--})$ as:
\begin{equation}
\psi_{V}(\hat{q})\approx
M\left[\slashed{\epsilon}+\frac{(\omega_{1}+\omega_{2})}{(m_{2}\omega_{1}+m_{1}\omega_{2})}
\hat{q}.\epsilon\right]\chi_{1}(\hat{q})+\left[\slashed{\epsilon}\slashed{P}-
(\slashed{P}\hat{q}.\epsilon+\slashed{P}\slashed{\epsilon}\slashed{\hat{q}})
\frac{(\omega_{1}+\omega_{2})}{(m_{2}\omega_{1}+m_{1}\omega_{2})}\right]\chi_{2}(\hat{q})
\end{equation}
where, we have been able to express the instantaneous wave
function $\psi^{V}(\widehat{q})$ in terms of only the leading
Dirac structures associated with amplitudes $\chi_{1}$ and
$\chi_{2}$. Putting the wave function Eq.(21) into the first two
Salpeter equations of Eq.(9) and by evaluating trace over the
$\gamma$-matrices on both sides, we can obtain two independent BS
coupled integral equations:
\begin{eqnarray}
 &&\nonumber (M-\omega_{1}-\omega_{2})(N_{1}\chi_{1}
+N_{2}\chi_{2})=-\int\frac{d^{3}\hat{q}'}{(2\pi)^{3}}K(\hat{q},\hat{q}')
[N'_{1}\chi'_{1} +N'_{2}\chi'_{2}]\\&&
 (M+\omega_{1}+\omega_{2})(N_{1}\chi_{1}
-N_{2}\chi_{2})=\int\frac{d^{3}\hat{q}'}{(2\pi)^{3}}K(\hat{q},\hat{q}')
[N'_{1}\chi'_{1}-N'_{2}\chi'_{2}]
\end{eqnarray}
where
\begin{eqnarray}
 &&\nonumber N_{1}=
 \left[\omega_{1}\omega_{2}A+m_{1}+m_{2}- A\hat{q}^{2}\right]\hat{q}.\epsilon\\&&
  \nonumber N_{2}=\left[2A\omega_{1}m_{2}-\omega_{1}-2A\omega_{2}m_{1}-m_{1}m_{2}A-\omega_{2}\right]\hat{q}.\epsilon\\&&
  \nonumber N'_{1}=(\omega_{1}\omega_{2}A'-m_{1}m_{2}A'-A'\hat{q}^{2})\hat{q}'.\epsilon-(m_{1}+m_{2})\hat{q}.\epsilon\\&&
  \nonumber N'_{2}=(2A'\omega_{1}m_{2}-2A'\omega_{2}m_{1})\hat{q}'.\epsilon-(\omega_{1}-\omega_{2})\hat{q}.\epsilon\\&&
   \nonumber A=\frac{\omega_{1}+\omega_{2}}{\omega_{1}m_{2}+\omega_{2}m_{1}}\\&&
  \nonumber A'=\frac{\omega'_{1}+\omega'_{2}}{\omega'_{1}m_{2}+\omega'_{2}m_{1}}\\&&
  \nonumber \omega_{1,2}^{2} = m_{1,2}^{2} + \hat{q}^{2}\\&&
  \omega'^{2}_{1,2}= m_{1,2}^{2} + \hat{q}'^{2}
\end{eqnarray}

These above two equations also reduce to the corresponding coupled
equations for vector quarkonia in \cite{hluf16} under equal mass
condition. For the solution of Eq.(16) and Eq.(22), we briefly
mention some features of the BS formulation employed. We now
introduce the BS kernel, $K(q,q')$ \cite{mitra01,bhatnagar14},
which is taken to be one-gluon-exchange like as regards the spin
dependence $(\gamma_{\mu}\bigotimes\gamma^{\mu})$, and color
dependence,
$(\frac{1}{2}\vec{\lambda}_{1}.\frac{1}{2}\vec{\lambda}_{2})$, and
has a scalar part $V$:

\begin{eqnarray}
&&\nonumber K(q,q') =
(\frac{1}{2}\vec{\lambda}_{1}.\frac{1}{2}\vec{\lambda}_{2})(\gamma_{\mu}\otimes\gamma^{\mu})V(q-q')\\&&
\nonumber\ V(\hat{q},\hat{q}')=
\frac{3}{4}\omega^{2}_{q\bar{q}}\int
d^{3}\vec{r}\left[r^{2}(1+4\hat{m}_{1}\hat{m}_{2}A_{0}M_{>}^{2}r^{2})^{-\frac{1}{2}}-\frac{C_{0}}{\omega_{0}^{2}}\right]
e^{i(\hat{q}-\hat{q}').\vec{r}}\\&&
\nonumber\omega^{2}_{q\bar{q}}=4M_{>}\hat{m}_{1}\hat{m}_{2}\omega^{2}_{0}\alpha_{s}(M_{>}
^{2})\\&&
\nonumber\alpha_{s}(M_{>}^{2})=\frac{12\pi}{33-2n_{f}}\left[log(\frac{M_{>}^{2}}{\wedge^{2}})\right]^{-1}\\&&
\nonumber
\hat{m}_{1,2}=\frac{1}{2}\left[1\pm\frac{(m^{2}_{1}-m^{2}_{2})}{M_{>}^{2}}\right]\\&&
\nonumber
\kappa=(1+4\hat{m}_{1}\hat{m}_{2}A_{0}M_{>}^{2}r^{2})^{-\frac{1}{2}}\\&&
\ M_{>}= Max(M, m_{1}+m_{2}).
\end{eqnarray}
We can write the complete potential as in \cite{hluf16}:
\begin{eqnarray}
&&\nonumber\
K(\hat{q},\hat{q}')=\overline{V}(\hat{q})\delta^{3}(\hat{q}-\hat{q}')\\&&
\nonumber
\overline{V}(\hat{q})=\omega^{2}_{q\bar{q}}\left[\kappa\overrightarrow{\nabla}^{2}_{\hat{q}}
+\frac{C_{0}} {\omega^{2}_{0}}\right](2\pi)^{3}\\&&
\kappa=\left(1-4\hat{m}_{1}\hat{m}_{2}A_{0}M^{2}\overrightarrow{\nabla}^{2}_{\hat{q}}\right)^{-1/2}
\end{eqnarray}
The potential is purely confining (as in Ref.
\cite{koll,hluf16,babutsidze}, and the Martin potential
\cite{olsson} employed for heavy mesons). Here in the expression
for $K(\widehat{q},\widehat{q}')$, the proportionality of
$\omega_{q\overline{q}}^{2}$ is needed to provide a more direct
QCD motivation \cite{mitra01} to confinement and the constant term
$C_{0}/\omega_{0}^{2}$ is designed to take account of the correct
zero point energies. Hence the term
$(1-4\hat{m}_{1}\hat{m}_{2}A_{0}M^{2}\overrightarrow{\nabla}^{2}_{\hat{q}})^{-1/2}$
in the above expression is responsible for effecting a smooth
transition from harmonic ($q\overline{q}$) to linear
($Q\overline{Q}$) confinement.
\bigskip

To decouple the mass spectral equations Eq.(16) and Eq.(22), we
first add them and subtract the second equations from the first
equations. For a kernel expressed as
$K(\widehat{q}-\widehat{q}')\approx\overline{V}(\widehat{q})\delta^{3}(\widehat{q}-\widehat{q}')$,
we get two algebraic equations which are still coupled. Then from
one of the two equations so obtained, we eliminate
$\phi_{1}(\hat{q})$ and $\chi_{1}(\hat{q})$ in terms of
$\phi_{2}(\hat{q})$ and $\chi_{2}(\hat{q})$, and plug these
expressions for $\phi_{1}(\hat{q})$ and $\chi_{1}(\hat{q})$ in the
second equations of the coupled set so obtained to get a decoupled
equations in $\phi_{2}(\hat{q})$ and $\chi_{2}(\hat{q})$
respectively from Eq.(16) and Eq.(22). Similarly, we eliminate
$\phi_{2}(\hat{q})$ and $\chi_{2}(\hat{q})$ from the second
equation of the set of coupled algebraic equations in terms of
$\phi_{1}(\hat{q})$ and $\chi_{1}(\hat{q})$, and plug these
expressions into the first two equations to get a decoupled
equations entirely in $\phi_{1}(\hat{q})$ and $\chi_{1}(\hat{q})$,
respectively from Eq.(16) and Eq.(22). Thus, we get two identical
decoupled equations, one entirely in $\phi_{1}(\hat{q})$ and
$\chi_{1}(\hat{q})$, and the other that is entirely in,
$\phi_{2}(\hat{q})$ and $\chi_{2}(\hat{q})$ respectively for heavy
pseudoscalar and vector mesons. Employing the limit,
$\omega_{1,2}\approx m_{1,2}$, the mass spectral equations can be
expressed for heavy pseudoscalar mesons as:
\begin{eqnarray}
 &&\nonumber \left[\frac{M^{2}}{4}-\frac{1}{4}(m_{1}+m_{2})^{2}-\hat{q}^{2}\right]\phi_{1}(\hat{q})=
 -2(m_{1}+m_{2})\omega^{2}_{q\bar{q}}\left[\kappa\overrightarrow{\nabla}^{2}_{\hat{q}}
+\frac{C_{0}} {\omega^{2}_{0}}\right]\phi_{1}(\hat{q})-
 4\omega^{4}_{q\bar{q}}\left[\kappa\overrightarrow{\nabla}^{2}_{\hat{q}}
+\frac{C_{0}} {\omega^{2}_{0}}\right]^{2}\phi_{1}(\hat{q})\\&&
 \left[\frac{M^{2}}{4}-\frac{1}{4}(m_{1}+m_{2})^{2}-\hat{q}^{2}
 \right]\phi_{2}(\hat{q})=-2(m_{1}+m_{2})\omega^{2}_{q\bar{q}}\left[\kappa\overrightarrow{\nabla}^{2}_{\hat{q}}
+\frac{C_{0}}
{\omega^{2}_{0}}\right]\phi_{2}(\hat{q})-4\omega^{4}_{q\bar{q}}\left[\kappa\overrightarrow{\nabla}^{2}_{\hat{q}}
+\frac{C_{0}} {\omega^{2}_{0}}\right]^{2}\phi_{2}(\hat{q})
\end{eqnarray}
and for heavy vector mesons as:
\begin{eqnarray}
 &&\nonumber \left[\frac{M^{2}}{4}-\frac{1}{4}(m_{1}+m_{2})^{2}-\hat{q}^{2}\right]\chi_{1}(\hat{q})=
 -(m_{1}+m_{2})\omega^{2}_{q\bar{q}}\left[\kappa\overrightarrow{\nabla}^{2}_{\hat{q}}
+\frac{C_{0}} {\omega^{2}_{0}}\right]\chi_{1}(\hat{q})+
 2\omega^{4}_{q\bar{q}}\left[\kappa\overrightarrow{\nabla}^{2}_{\hat{q}}
+\frac{C_{0}} {\omega^{2}_{0}}\right]^{2}\chi_{1}(\hat{q})\\&&
 \left[\frac{M^{2}}{4}-\frac{1}{4}(m_{1}+m_{2})^{2}-\hat{q}^{2}
 \right]\chi_{2}(\hat{q})=-(m_{1}+m_{2})\omega^{2}_{q\bar{q}}\left[\kappa\overrightarrow{\nabla}^{2}_{\hat{q}}
+\frac{C_{0}}
{\omega^{2}_{0}}\right]\chi_{2}(\hat{q})+2\omega^{4}_{q\bar{q}}\left[\kappa\overrightarrow{\nabla}^{2}_{\hat{q}}
+\frac{C_{0}} {\omega^{2}_{0}}\right]^{2}\chi_{2}(\hat{q})
\end{eqnarray}
These two decoupled Eq. (26) and Eq. (27), would resemble harmonic
oscillator equations, except for second term on the RHS of these
equations. It will be shown later from the definition of kernel in
Eq.(24), that these second terms are negligible in comparison to
the first terms on the RHS, and can be dropped, and these
equations would then resemble exact harmonic oscillator equations.
However, due to identical nature of these equations, their
solutions are written as:
$\phi_{1}(\hat{q})=\phi_{2}(\hat{q})\approx\phi_{P}(\hat{q})$ for
pseudoscalar mesons and
$\chi_{1}(\hat{q})=\chi_{2}(\hat{q})\approx \phi_{V}(\hat{q})$ for
vector mesons, which represent the eigenfunctions of unequal mass
heavy pseudoscalar and vector mesons obtained by solving the full
Salpeter equation and we can write the wave function for heavy
pseudoscalar mesons as:
\begin{equation}
\psi_{P}(\hat{q}) \approx
\left[M+\slashed{P}+\frac{\slashed{\hat{q}}M(\omega_{2}-\omega_{1})}{m_{2}\omega_{1}+m_{1}\omega_{2}}
+\frac{\slashed{\hat{q}}\slashed{P}(\omega_{2}+\omega_{1})}{m_{2}\omega_{1}+m_{1}\omega_{2}}
\right]\gamma_{5}\phi_{P}(\hat{q})
\end{equation}
and the wave function for heavy vector mesons as:
\begin{equation}
\psi_{V}(\hat{q})\approx
\left[M(\slashed{\epsilon}+\frac{(\omega_{1}+\omega_{2})}{(m_{2}\omega_{1}+m_{1}\omega_{2})}
\hat{q}.\epsilon)+\slashed{\epsilon}\slashed{P}-
(\slashed{P}\hat{q}.\epsilon+\slashed{P}\slashed{\epsilon}\slashed{\hat{q}})
\frac{(\omega_{1}+\omega_{2})}{(m_{2}\omega_{1}+m_{1}\omega_{2})}\right]\phi_{V}(\hat{q})
\end{equation}
\end{enumerate}
As mentioned above, we have dropped the second terms of Eq. (26)
and Eq. (27), in comparison to the first terms on the RHS of Eq.
(26) and Eq. (27). As the coefficients,
$\rho_{2}=4\omega^{4}_{q\bar{q}}$ and
$\rho'_{2}=2\omega^{4}_{q\bar{q}}$, respectively for heavy
pseudoscalar and vector mesons, associated with second term have a
very small contribution $(\leq 1.2556\%)$ in comparison to the
coefficient $\rho_{1}=2(m_{1}+m_{2})\omega^{2}_{q\bar{q}}$ and
$\rho'_{1}=(m_{1}+m_{2})\omega^{2}_{q\bar{q}}$ respectively for
heavy pseudoscalar and vector mesons, associated with first term,
due to $\omega^{4}_{q\bar{q}}<< \omega^{2}_{q\bar{q}}$ for
$\eta_{c}$, $B_{c}$, $\eta_{b}$, $J/\psi$, $B^{*}_{c}$ and
$\Upsilon$). The numerical values of these coefficients, and their
percentage ratio for heavy pseudoscalar and vector mesons are
given in Table 1 below, which justifies this term being dropped or
negligible. We left the signs and the units with equations of Eq.
(26) and Eq. (27), we have taken only the constant coefficients
when we calculate the contributions of the first and second terms
of these equations.
\begin{table}[htbp]
 \renewcommand{\tabcolsep}{30.6pt}
   \begin{center}
\begin{tabular}{cccc}
  \hline
     &$\rho_{1}$ or $\rho'_{1}$&$\rho_{2}$ or $\rho'_{2}$&$\frac{\rho_{2}}{\rho_{1}}\%$ or $\frac{\rho'_{2}}{\rho'_{1}}\%$\\\hline
    $c\overline{c}$&0.1115&0.0014&1.2556\% \\
    $c\overline{b}$&0.3188&0.0024&0.7528\% \\
     $b\overline{b}$&0.9653&0.0091&0.9427\% \\\hline
     \end{tabular}
\end{center}
\caption{Numerical values of coefficients for heavy pseudoscalar
and vector mesons and their percentage ratio,
  along with the corresponding values. The input parameters of
our model are: $C_{0}=0.21$, $\omega_{0}=.15GeV.$, QCD length
scale $\Lambda=0.200GeV.$, $A_{0}=0.01$, and the input quark
masses, $m_{c}=1.49GeV.$, and $m_{b}=5.070GeV.$}
 \end{table}

The form of $K(\hat{q},\hat{q}')$ in Eq.(24) suggests that Eqs.
(26) and Eqs. (27) have to be solved numerically. However, to
solve these equations analytically, we follow an analytical
procedure on lines of Ref.\cite{hluf15,hluf16}, where we treat
$\kappa$ as a "correction" factor due to small value of parameter,
$A_{0}$ ($A_{0}<< 1$), while we work in an approximate harmonic
oscillator basis, due to its transparency in bringing out the
dependence of mass spectral equations on the total quantum number
$N$. (In this connection, we wish to mention that recently,
harmonic oscillator basis has also been widely employed to study
heavy mesons using a Light-front quark model \cite{peng12}.) The
latter is achieved through the effective replacement,
$\kappa=(1-4\hat{m}_{1}\hat{m}_{2}A_{0}M^{2}\overrightarrow{\nabla}_{\hat{q}}^{2})^{-\frac{1}{2}}\Rightarrow
(1+8\hat{m}_{1}\hat{m}_{2}A_{0}(N+\frac{3}{2}))^{-\frac{1}{2}}$,
(which is quite valid for heavy $c\overline{c}$, $c\overline{b}$
and $b\overline{b}$ systems). With this, we can reduce Eq. (26)
and Eq. (27) to equations of a simple quantum mechanical 3D-
harmonic oscillator with coefficients depending on the hadron mass
$M$ and total quantum number $N$. The wave function satisfies the
3D BSE for heavy pseudoscalar and vector mesons respectively as
given below:
 \begin{equation}
 \left[\frac{M^{2}}{4}-\frac{1}{4}(m_{1}+m_{2})^{2}-\hat{q}^{2}
 \right]\phi_{P}(\hat{q})=-\beta^{4}_{P}
 \left[\overrightarrow{\nabla}^{2}_{\hat{q}}+\frac{C_{0}}
{\omega^{2}_{0}}\sqrt{1+8\hat{m}_{1}\hat{m}_{2}A_{0}(N+\frac{3}{2})}\right]\phi_{P}(\hat{q})
\end{equation}
and
\begin{equation}
 \left[\frac{M^{2}}{4}-\frac{1}{4}(m_{1}+m_{2})^{2}-\hat{q}^{2}
 \right]\phi_{V}(\hat{q})=-\beta^{4}_{V}
 \left[\overrightarrow{\nabla}^{2}_{\hat{q}}+\frac{C_{0}}
{\omega^{2}_{0}}\sqrt{1+8\hat{m}_{1}\hat{m}_{2}A_{0}(N+\frac{3}{2})}\right]\phi_{V}(\hat{q})
\end{equation}
where the inverse range parameter for heavy pseudoscalar and
vector mesons respectively is,
$\beta_{P}=\left[2\frac{(m_{1}+m_{2})\omega^{2}_{q\bar{q}}}{\sqrt{1+8\hat{m}_{1}\hat{m}_{2}A_{0}(N+\frac{3}{2})}}\right]^{\frac{1}{4}}$
and
$\beta_{V}=\left[\frac{(m_{1}+m_{2})\omega^{2}_{q\bar{q}}}{\sqrt{1+8\hat{m}_{1}\hat{m}_{2}A_{0}(N+\frac{3}{2})}}\right]^{\frac{1}{4}}$
and is dependent on the input kernel parameters and contains the
dynamical information. The numerical values of inverse range
parameters $\beta_{P}$, and $\beta_{V}$ for various heavy
pseudoscalar and vector mesons in the mass spectrum studied in
this paper are listed in the Table 2 below.
\begin{table}[h]
\begin{center}
\renewcommand{\tabcolsep}{25pt}
\begin{tabular}{llllll}
  \hline
    &$1S$&$2S$&$1D$&$3S$&$2D$\\\hline
    $\beta_{\eta_{c}}$&0.3314&0.3282&&0.3252&\\
    $\beta_{B_{c}}$&0.5588&0.5516&&0.5447&\\
    $\beta_{\eta_{b}}$&0.9753&0.9660&&0.9572&\\
    $\beta_{J/\psi}$&0.2343&0.2321&0.2321&0.2300&0.2300\\
    $\beta_{B^{*}_{c}}$&0.3952&0.3900&0.3900&0.3852&0.3852\\
    $\beta_{\Upsilon}$&0.6896&0.6831&0.6831&0.6768&0.6768\\
    \hline
   \end{tabular}
   \end{center}
   \caption{$\beta_{P}$ and $\beta_{V}$ values for ground
state and excited states of $\eta_{c}$, $B_{c}$, $\eta_{b}$,
$J/\psi$, $B^{*}_{c}$ and $\Upsilon$ (in GeV units) in present
calculation (BSE-CIA).}
\end{table}
Solutions of Eq.(30) and Eq.(31), their eigenfunctions and
eigenvalues are well known. The analytical solutions of these
equations for ground and excited states are given below.
\begin{eqnarray}
&&\nonumber
\phi_{P(V)}(1S,\hat{q})=\frac{1}{\pi^{3/4}\beta^{3/2}}e^{-\frac{\hat{q}^{2}}{2\beta^{2}}}\\&&
\nonumber \phi_{P(V)}(2S,\hat{q})=
(\frac{3}{2})^{1/2}\frac{1}{\pi^{3/4}\beta^{3/2}}(1-\frac{2\hat{q}^{2}}{3\beta^{2}})e^{-\frac{\hat{q}^{2}}{2\beta^{2}}}\\&&
\nonumber
\phi_{V}(1D,\hat{q})=(\frac{4}{15})^{1/2}\frac{1}{\pi^{3/4}\beta^{7/2}}\widehat{q}^{2}e^{-\frac{\hat{q}^{2}}{2\beta^{2}}}\\&&
\nonumber
\phi_{P(V)}(3S,\hat{q})=(\frac{15}{8})^{1/2}\frac{1}{\pi^{3/4}\beta^{3/2}}
(1-\frac{20\hat{q}^{2}}{15\beta^{2}}+\frac{4\hat{q}^{4}}{15\beta^{4}})e^{-\frac{\hat{q}^{2}}{2\beta^{2}}}\\&&
\nonumber\phi_{V}(2D,\hat{q})=(\frac{14}{15})^{1/2}\frac{1}{\pi^{3/4}\beta^{7/2}}
(1-\frac{2\widehat{q}^{2}}{7\beta^{2}}) \widehat{q}^{2}
e^{-\frac{\hat{q}^{2}}{2\beta^{2}}}\\&&
\end{eqnarray}
where $\beta=\beta_{P(V)}$. We will use the wave functions for a
description of various transition amplitudes of these heavy
pseudoscalar and vector mesons in our future work. We now give the
plots of these normalized wave functions Vs. $\widehat{q}$ (in
Gev.) for different states of heavy pseudoscalar and vector mesons
( such as composite of $c\overline{c}$, $c\overline{b}$ and
$b\overline{b}$) in Fig.1 and in Fig.2, respectively. It can be
seen from these plots that the wave functions corresponding to
$nS$ and $nD$ states have $n-1$ nodes.
\begin{figure}[h]
\centering
\includegraphics[width=6cm]{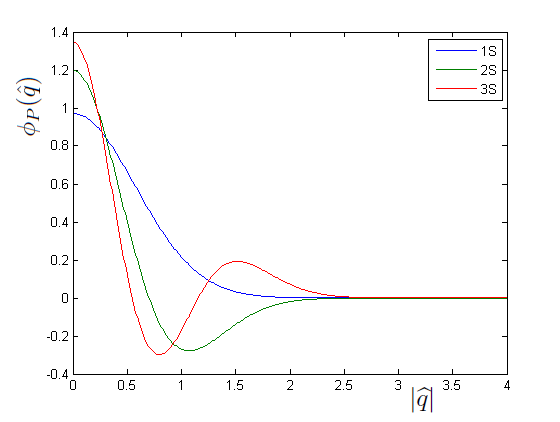}
\includegraphics[width=6cm]{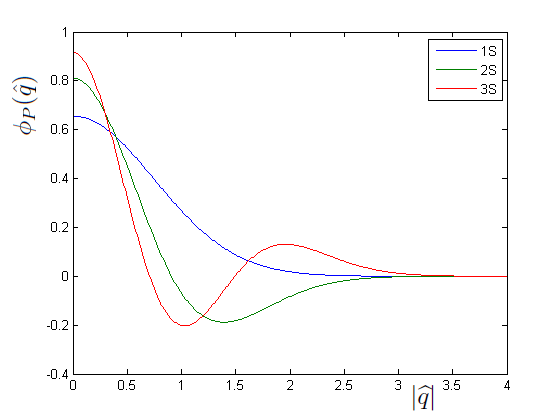}\\
\includegraphics[width=6cm]{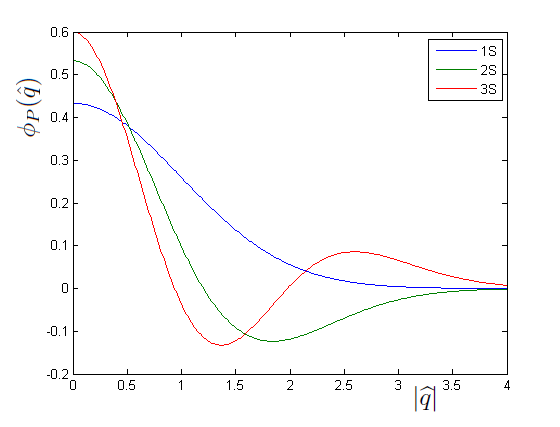}
\caption{Plots of wave functions for states $(1S,...,3S)$ Vs
$\widehat{q}$ (in Gev.) for pseudoscalar mesons, such as;
$\eta_{c}$, $B_{c}$ and $\eta_{b}$, respectively.}
\end{figure}
\begin{figure}[h]
\centering
\includegraphics[width=6cm]{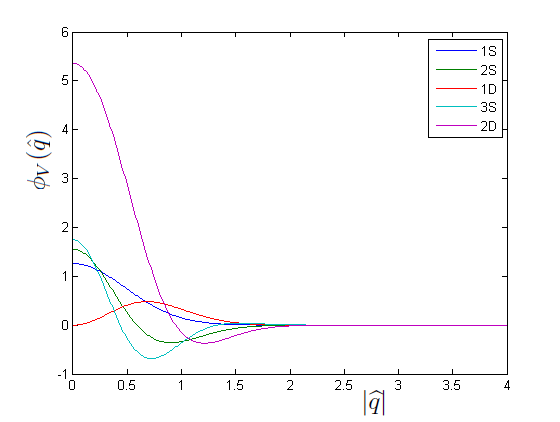}
\includegraphics[width=6cm]{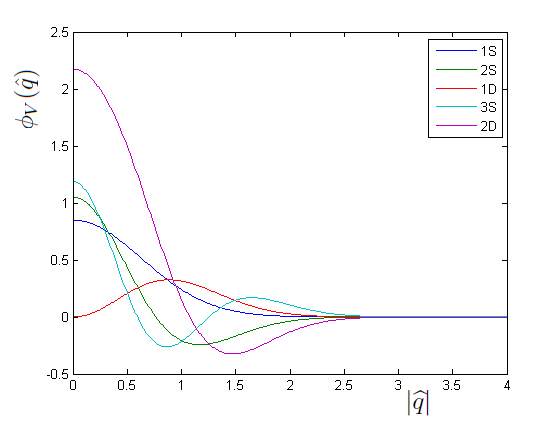}\\
\includegraphics[width=6cm]{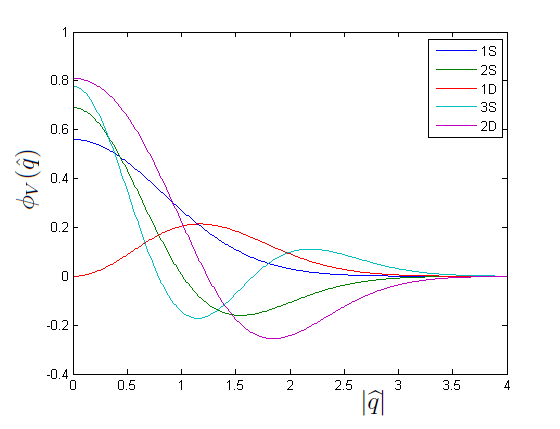}
\caption{Plots of wave functions for states $(1S,...,2D)$ Vs
$\widehat{q}$ (in Gev.) for vector mesons, such as; $J/\psi$,
$B^{*}_{c}$ and $\Upsilon$), respectively.}
\end{figure}

The mass spectrum equation of ground and excited states for heavy
pseudoscalar and vector mesons are written respectively as:
\begin{equation}
 \frac{1}{2\beta_{P}^{2}}\left[\frac{M^{2}}{4}-\frac{1}{4}(m_{1}+m_{2})^{2}+\frac{C_{0}\beta_{P}^{4}}{\omega^{2}_{0}}
\sqrt{1+8\hat{m}_{1}\hat{m}_{2}A_{0}(N+\frac{3}{2})}\right]=N+\frac{3}{2};
N=0,2,4,...
\end{equation}
\begin{equation}
 \frac{1}{2\beta_{V}^{2}}\left[\frac{M^{2}}{4}-\frac{1}{4}(m_{1}+m_{2})^{2}+\frac{C_{0}\beta_{V}^{4}}{\omega^{2}_{0}}
\sqrt{1+8\hat{m}_{1}\hat{m}_{2}A_{0}(N+\frac{3}{2})}\right]=N+\frac{3}{2};
N=0,2,4,...
\end{equation}
where $N=2n+l=0,2,4,...$ for $n=0, 1, 2, ..$ and the orbital
quantum
number $l=0$ for S-state and $l=2$ for D-state.\\
\textbf{Numerical Results}\\
The input parameters of our model are: $C_{0}=0.21$,
$\omega_{0}=.15GeV$, QCD length scale $\Lambda=0.200GeV$,
$A_{0}=0.01$, and the input quark masses, $m_{c}=1.49GeV$, and
$m_{b}=5.070GeV$. The results of mass spectral predictions of
heavy pseudoscalar and vector mesons for both ground and
 excited states with the above set of parameters are given in
tables 3 and 4.
\begin{table}[htbp]
\renewcommand{\tabcolsep}{16.6pt}
\centering
   \begin{center}
   \begin{tabular}{cccccc}
  \hline
             State&BSE-CIA&Expt.\cite{Olive14}&PM\cite{rai13}&Rel. PM\cite{ebert02}&Rel. QM\cite{god04}\\\hline
             $M_{\eta_{c}(1S)}$&2.9489&2.983$\pm$0.0007&2.985&2.980&\\
             $M_{\eta_{c}(2S)}$&3.7296&3.639$\pm$0.0013&3.626&3.589&\\
             $M_{\eta_{c}(3S)}$&4.3623&                &4.047&     &\\\hline
             $M_{B_{c}(1S)}$&6.1611&6.2756$\pm0.0011$  &6.256&6.270&6.271\\
             $M_{B_{c}(2S)}$&6.8442&                   &6.929&6.835&6.855\\
             $M_{B_{c}(3S)}$&7.4501&                   &7.308&7.193&7250\\\hline
             $M_{\eta_{b}(1S)}$&8.8590&9.398 $\pm$0.0032&9.425&9.314&\\
             $M_{\eta_{b}(2S)}$&9.6866&9.999$\pm$0.0028&10.012&9.931&\\
             $M_{\eta_{b}(3S)}$&10.4354&               &10.319&10.288&\\
             \hline
      \end{tabular}
   \end{center}
   \caption{Mass spectrum for ground and excited states of heavy pseudoscalar mesons (in GeV units)
in present calculation (BSE-CIA).}
    \end{table}\\

\begin{table}[htbp]
\renewcommand{\tabcolsep}{16.6pt}
\centering
   \begin{center}
   \begin{tabular}{cccccc}
  \hline
             State&BSE-CIA&Expt.\cite{Olive14}&PM\cite{rai13}&Rel. PM\cite{ebert02}&Rel. QM\cite{god04}\\\hline
             $M_{J/\psi(1S)}$&3.1003&3.0969$\pm$ 0.000011&3.096&3.097&\\
             $M_{\psi(2S)}$&3.6468  &3.6861$\pm$ 0.00034&3.690&3.686&\\
             $M_{\psi(1D)}$&3.6468  &3.773$\pm$ 0.00033&      &3.814&\\
             $M_{\psi(3S)}$&4.1134  &4.03$\pm$ 0.001&4.082&     &\\
             $M_{\psi(2D)}$&3.1134  &4.191$\pm$0.005&            &     &\\\hline
             $M_{B^{*}_{c}(1S)}$&6.4718&            &6.314&6.332&6.338\\
             $M_{B^{*}_{c}(2S)}$&6.9381&            &6.968&6.881&6.887\\
             $M_{B^{*}_{c}(1D)}$&6.9381&            &            &7.072&7.028\\
             $M_{B^{*}_{c}(3S)}$&7.3643&            &7.326 &7.235&7.272\\
             $M_{B^{*}_{c}(2D)}$&7.3643&            &            &     &7.365\\\hline
             $M_{\Upsilon(1S)}$&9.6476&9.4603$\pm$ 0.00026&9.461&9.460&\\
             $M_{\Upsilon(2S)}$&10.1944&10.0233$\pm$0.00031&10.027&10.023&\\
             $M_{\Upsilon(1D)}$&10.1944&                   &     &10.147&\\
             $M_{\Upsilon(3S)}$&10.7043&10.3552$\pm$0.00005&10.329&10.355&\\
             $M_{\Upsilon(2D)}$&10.7043&                   &     &10.442&\\
             \\\hline
      \end{tabular}
   \end{center}
   \caption{Mass spectrum for ground
and excited states of heavy vector mesons (in GeV units) in
present calculation (BSE-CIA).}
    \end{table}
\section{Discussions and Conclusion}
In this work, using BSE framework under CIA for heavy quarks, we
have done the analytical calculation of the mass spectrum of
ground states and excited states of heavy quarks pseudoscalar and
vector mesons in approximate harmonic oscillator basis. This work
is a generalization of our previous work \cite{hluf16} involving
equal mass quarkonia to calculation of mass spectra of unequal
mass quarkonia studied in this paper. Further, it can be easily
checked that our equations right from the coupled equations for
unequal mass pseudoscalar mesons (Eqs.(16-17)) and unequal mass
vector mesons (Eqs.(22-23)), all the way up to their mass spectral
equations (Eqs.(33-34)) reduce to the corresponding equations in
\cite{hluf16} for equal mass quarkonia when two quarks are taken
to have the same mass, validating the correctness of our
calculations.

It is to be noted that the coupled integral equations one obtains
by solving the 4X4 BSE for unequal mass mesons are quite
complicated, which in most works is solved only numerically
\cite{wang10}, but in this study, we have attempted to solve the
coupled integral equations analytically, and we have explicitly
shown that in the heavy quark approximation (valid for quarkonium
systems), these equations can be decoupled, and finally analytical
solutions (both mass spectrum and eigenfunctions) of these
equation can be obtained using approximate harmonic oscillator
basis.

All numerical calculations have been done using Math lab. We first
fit our parameters to the ground state masses of $\eta_{c}$,
$B_{c}$, $\eta_{b}$, $J/\psi$, $B^{*}_{c}$ and $\Upsilon$ with
data. Using the input parameters, we obtained the best fit to
these ground state masses. With the same set of parameters, we
calculate the masses of all the other excited states of
$\eta_{c}$, $B_{c}$, $\eta_{b}$, $J/\psi$, $B^{*}_{c}$ and
$\Upsilon$. These input parameters of our model are the same with
our recent papers\cite{hluf16,hluf17}. The calculated mass
spectrum in the framework of BSE are listed in Tables 3 and 4. The
results obtained for masses of ground and excited states of
$\eta_{c}$, $B_{c}$, $\eta_{b}$, $J/\psi$, $B^{*}_{c}$ and
$\Upsilon$ are in reasonable agreement with experiment. However, a
wide range of variation in masses of various states can be seen in
Tables 3 and 4 and the experimental data on masses of some of
these states is not yet currently available.

From the tables, one can read out for vector mesons, there is a
degeneracy in the masses of $S$ and $D$ states with the same
principal quantum number $N$ for $J/\Psi$, $B^{*}_{c}$ and
$\Upsilon$. This is due to the fact that we did not incorporate
the one-gluon-exchange (OGE) effects in the kernel, and used only
the confining part of interaction taking analogy from
\cite{koll,hluf15,hluf16,babutsidze,olsson} for heavy mesons.
However our results also reflect the fact that the OGE term
becomes more and more important for very heavy mesons and the
inclusion of OGE terms in the potential will lift up the
degeneracy in these
states.\\

However, our main focus was to show that this problem of $4\times
4$ BSE under heavy quark approximation can indeed be handled
analytically for both masses, as well as the wave functions in an
approximate harmonic oscillator basis. The analytical forms of
wave functions are obtained as solutions of mass spectral
equations (that are derived from 3D BSE). Analytical forms of the
wave functions for both $S$ and $D$ states for $n=0,1,2,...$ for
$c\overline{c}$, $c\overline{b}$ and $b\overline{b}$ systems thus
obtained are given in Eq.(32). We have also plotted the graphs of
all these wave functions for the states, $1S,...,3S$ for
$\eta_{c}$, $B_{c}$ and $\eta_{b}$ mesons, and for the states
$1S,2S,1D,3S$ and $2D$ for $J/\psi$, $B^{*}_{c}$ and $\Upsilon$
mesons, in Fig. 1 and Fig. 2.

We are not aware of any other BSE framework, involving $4\times 4$
BS amplitude, that treats the mass spectral problem involving
heavy mesons analytically. To the best of our knowledge, all the
other $4\times4$ BSE approaches treat this problem numerically
just after they obtain the coupled set of equations (see
Ref.\cite{wang10}). We further wish to mention that the over all
features of our plots of wave functions in Eqs.(32) derived in our
framework are very similar to the corresponding plots of wave
functions obtained by purely numerical methods in \cite{wang10},
suggesting that our approach is not only in good agreement with
the numerical approaches followed in other works, but also gives a
deeper understanding of the problem, by showing an explicit
dependence of the mass spectrum on principal quantum number $N$.
In the present paper, we have restricted our study only to the
analytical calculation of heavy pseudoscalar and vector mesons
mass spectrum and their corresponding wave functions. We will
extend this study to the calculation of various transition
amplitudes of these heavy pseudoscalar and vector mesons in our
framework to do as our further work. It is seen that in this
framework, we were able to obtain fairly good values of masses for
their ground as well as their excited states of $\eta_{c}$,
$B_{c}$,
$\eta_{b}$, $J/\psi$, $B^{*}_{c}$ and $\Upsilon$.\\

\textbf{Acknowledgements}:\\
 A good part of this work was carried out at Samara University, Ethiopia, and a part at Chandigarh
 University, India.  We thank these Institutions for the facilities provided during the course of
 this work.

\end{document}